\documentclass[conference]{IEEEtran}

\makeatletter
\newcommand{\newlineauthors}{%
  \end{@IEEEauthorhalign}\hfill\mbox{}\par
  \mbox{}\hfill\begin{@IEEEauthorhalign}
}
\makeatother

\IEEEoverridecommandlockouts
\usepackage[utf8]{inputenc}
\usepackage{alphabeta} % for displaying greek letters in the code snippets
\usepackage{cite}
\usepackage{amsmath,amssymb,amsfonts}
\usepackage{algorithmic}
\usepackage{graphicx}
\usepackage{textcomp}
\usepackage{xcolor}
\usepackage[frozencache=true]{minted}
\usepackage{physics}
\usepackage[hyphens]{url} % enabling linebreaks in references
\usepackage{hyperref}

\usepackage{tikz}
\usetikzlibrary{positioning}
\usetikzlibrary{calc}
\usetikzlibrary{shapes.geometric}

\usepackage[T1]{fontenc}
\usepackage{lmodern}
\setminted{fontsize=\footnotesize, fontfamily=courier, frame=lines}
\def\BibTeX{{\rm B\kern-.05em{\sc i\kern-.025em b}\kern-.08em
    T\kern-.1667em\lower.7ex\hbox{E}\kern-.125emX}}

\begin{document}

\title{An Open Source Software Stack for Tuning the Dynamical Behavior of Complex Power Systems}

\author{
\IEEEauthorblockN{Anna Büttner\IEEEauthorrefmark{1}\IEEEauthorrefmark{2},
Hans Würfel\IEEEauthorrefmark{1}\IEEEauthorrefmark{2}, Anton Plietzsch\IEEEauthorrefmark{2},
Michael Lindner\IEEEauthorrefmark{2}\IEEEauthorrefmark{3} and Frank Hellmann\IEEEauthorrefmark{2} \thanks{\IEEEauthorrefmark{1} These authors contributed equally.}}
\IEEEauthorblockA{\IEEEauthorrefmark{2} \textit{Potsdam Institute for Climate Impact Research}\\
 Telegrafenberg A31, 14473 Potsdam, Germany\\
Email: buettner@pik-potsdam.de, wuerfel@pik-potsdam.de}
\IEEEauthorblockA{\IEEEauthorrefmark{3}  \textit{Institute of Theoretical Physics, Technische Universit\"at Berlin} \\
 Hardenbergstr. 36, D-10623 Berlin, Germany}
 }

% for reference on how to edit author affiliations (see page 5f): https://ctan.mirror.norbert-ruehl.de/macros/latex/contrib/IEEEtran/IEEEtran_HOWTO.pdf 

\maketitle

\begin{abstract}

BlockSystems.jl and NetworkDynamics.jl are two novel software packages which facilitate highly efficient transient stability simulations of power networks. Users may specify inputs and power system design in a convenient modular and equation-based manner without compromising on speed or model detail.
Written in the high-level, high-performance programming language Julia\cite{Julia-2017} a rich open-source package ecosystem is available, which provides state-of-the-art solvers and machine learning algorithms\cite{SciML}.

Motivated by the recent interest in the Nordic inertia challenge \cite{inertia_challenge} we have implemented the Nordic5 test case \cite{hydro_and_wind} and tuned its control parameters by making use of the machine learning and automatic differentiation capabilities of our software stack.
\end{abstract}


\section{Introduction}
One of the main challenges in designing a simulation software for power system dynamics is the large amount of possible modeling choices at several levels of the power system.
At the bus level, the models range from algebraic constraints and simple swing equations up to high order machine models with complex governors and control loops.
At the grid level the systems of interest range from small toy models containing of a handful buses up to huge, nation-scale systems with complex topology and tens of thousands of buses with heterogeneous dynamics.
Finally, many different scenarios can be considered, e.g. voltage drops, line failures or power perturbations, to name just a few.

To handle this complexity it is often helpful to build models from the ground up, starting with simple component models and topologies and gradually increasing the complexity of the system. 
The hierarchical structure of power systems is beneficial for this kind of component-based modeling: both at the grid, as well as at the bus level, subsystems often have clearly defined interfaces.

\begin{figure}%[h]
\centering
\scalebox{0.6}{\begin{tikzpicture}
\node[draw,minimum width= 5cm, minimum height=0cm](sym) {\texttt{Symbolics.jl}};
\node[draw,minimum width= 6cm, right = .2cm of sym](diffeq) {\texttt{DifferentialEquations.jl}};
\node[draw,minimum width= 3cm, right = .2cm of diffeq](flux) {\texttt{Flux.jl}};

\node[draw, above = .2cm of sym, minimum width=4cm](mtk) {\texttt{ModelingToolkit.jl}};
\node[draw, above = .2cm of mtk, minimum width=3cm](blk) {\texttt{BlockSystems.jl}};

\node[draw, above = .2cm of $(diffeq.north)!0.25!(diffeq.north west)$, minimum width=4cm](nd) {\texttt{NetworkDynamics.jl}};
\node[draw, above = .2cm of $(flux.north west)$, minimum width=3.5cm](diffeqflux) {\texttt{DiffEqFlux.jl}};

\node[draw, above = .2cm of diffeqflux](pbt) {\texttt{ProBeTune.jl}};
\node[draw, dotted, above = .2cm of nd](pd) {\texttt{PowerDynamics.jl}};

\node[ellipse, draw, above = .6cm of $(blk.north west)!0.5!(pbt.north east)$, minimum width=3cm](nordic) {Nordic5 simulation};
\end{tikzpicture}}
  \caption{A diagram showing the relationships among the main packages in our software stack. \label{fig:softwarestack}}
\end{figure}
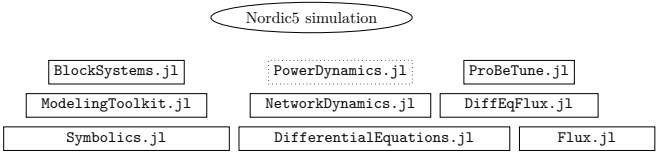

In this paper, we present a Julia software stack for modular, equation-based modeling of sub-nodal dynamics all the way up to machine learning-enabled parameter selection for controllers (see Fig.~\ref{fig:softwarestack}). Julia\cite{Julia-2017} is a high-level, high-performance programming language with a rich open-source package ecosystem. In particular, the Julia libraries for differential equation solving\cite{rackauckas2017differentialequations, diffeq_uebersicht} and "scientific machine learning"\cite{SciML, SciML_vergleich} have been shown to outperform competing solutions in other languages, both regarding speed and available features.

The \emph{two language problem} for simulation software, says that easy-to-use interfaces  are commonly written in high-level programming languages, whereas computational backends are implemented in efficient, but hard(er) to maintain low-level languages. Julia was designed to circumvent this problem by being both high-level and fast\cite{Julia-2017} (among other things due to just-in-time compilation).
Since our whole software stack is written in Julia there is no two language problem: Users can look through each step of modeling and computation without hitting obscure calls to underlying libraries written in low-level languages. 

Commercially available software is mostly closed source which is a problem in science. 
Those programs come with huge libraries of predefined building blocks and functionality which are hard to extend and have to be treated as black boxes in many cases.
Since the full stack is open source, the used models are exposed to the users and can be inspected, altered and extended to match the requirements.

In the next sections, we present a typical workflow for modeling and simulating a power system with the help of our software packages.
We start with \texttt{BlockSystems.jl}\footnote{\url{https://github.com/hexaeder/BlockSystems.jl}}, a frontend to model dynamics of power system components and control structures based on block diagrams and generate optimized Julia functions from the user input. These function are passed on to \texttt{NetworkDynamics.jl} \cite{lindner2021networkdynamics}, which combines user-supplied, modular node and edge dynamics, as well as a graph topology, into an efficient right hand side describing the whole system dynamics. The resulting differential equation can be solved with the state-of-the-art solver algorithms available in the \texttt{DifferentialEquations.jl} ecosystem. Finally, we demonstrate the machine learning and automatic differentiation capabilities of the SciML ecosystem, by tuning the parameters of the system such that its dynamics match those of a given specification, according to the recently proposed probabilistic tuning framework \cite{hellmann2021probetune}\footnote{\url{https://github.com/PIK-ICoNe/ProBeTune.jl}}.

\section{Software Packages}

\subsection{BlockSystems.jl}
\texttt{BlockSystems.jl} is a Julia package for modeling dynamical systems built from several subsystems (blocks) connected via an input-output relationship.
Each \emph{block} consists of a set of algebraic and differential equations as well as some meta information. The equations are symbolic representations based on the computer algebra system \texttt{Symbolics.jl} \cite{gowda2021high}.

The equations define the dynamics for all \emph{states} $\vb x$ of the block based on some \emph{parameters}. We define two types of states:
\begin{itemize}
\item outputs $o(t)$: time-dependent states, which are exposed to other blocks and
\item internal states $s(t)$: which are needed in order to calculate the outputs but are not exposed to other blocks.
\end{itemize}
The \emph{parameters} also fall in two categories:
\begin{itemize}
\item time dependent inputs $i(t)$ whose values are set externally and
\item time independent internal parameters $p$.
\end{itemize}

Using those symbols one can define three types of equations:
First-order ordinary differential equations for states $x \in\vb x$
\begin{equation}
    \dv{t} x(t) = f(\vb x, \vb i, \vb p, t)\,,
\end{equation}
where $\vb i$ is the set of all inputs and $\vb p$ is the set of all internal parameters;
explicit algebraic equations for state $x$ where the right hand side must not depend on $x$
\begin{equation}
  x(t) = f(\vb x\setminus\{x\}, \vb i, \vb p, t)\,;
\end{equation}
or implicit constraint equations
\begin{equation}\label{eq:implicit_constraints}
    0 = f(\vb x, \vb i, \vb p, t)\,.
\end{equation}

An example of such a block is the frequency dynamics of a simple swing node. 
In this model, the change of frequency $\omega$ depends on the balance between electrical power $P_e$ and mechanical power $P_m$. 
The equation is further parameterized by the inertia $M$ and the damping $D$:
\begin{equation}
    \dv{t}\omega(t) = \frac{1}{M}\bigl( P_m(t) - D\,\omega(t) - P_e(t) \bigr)
\end{equation}

This equation can be directly used to create an \texttt{IOBlock} with the inputs $P_m$ and $P_e$, the output $\omega$ and
the internal parameters $M$ and $D$.
%\newpage
\begin{minted}{julia}
@parameters t M D P_m(t) P_e(t)
@variables ω(t)
dt = Differential(t)

swing = IOBlock([dt(ω) ~ 1/M * (P_m - D*ω - P_e)],
                 [P_m, P_e], # inputs
                 [ω])        # outputs
\end{minted}

If there is a change in the electrical power demand $P_e$ the power difference has to be absorbed by the damping term. If $P_m$ is fixed the swing node will reach a new steady-state at a different frequency. With \texttt{BlockSystems.jl} we can easily combine blocks to create bigger systems. We will define a PID-controller to adapt $P_m$ after the change of demand in $P_e$ in order to restore the initial frequency. The value of the controller signal equals the sum of the input, the integral of the input, and the derivative of the input. For the output of the whole block, the controller signal gets subtracted from the fixed initial value of $P_m = 1$.

\begin{minted}{julia}
@parameters input(t)
@variables int(t) out(t) pid(t)
pid = IOBlock([dt(int) ~ input,
               pid ~ input + int + dt(input),
               out ~ 1 - pid],
              [input],
              [out])
\end{minted}

\begin{figure}[h]
\centering
\scalebox{0.65}{\begin{tikzpicture}
\node[draw,thick, text width=6cm](swing) {Swing Equation
\begin{equation*}
    \dv{t}\omega(t) = \frac 1 M \bigl(P_m(t) - D\,\omega - P_e(t) \bigr)
\end{equation*}};

\coordinate (pe) at ($(swing.north west)!0.5!(swing.west)$);
\coordinate (pm) at ($(swing.south west)!0.5!(swing.west)$);
\node[anchor=south west] at (swing.east){$\omega$};
\node[anchor=south east] at (pe){$P_e$};
\node[anchor=south east] at (pm){$P_m$};
\draw[<-, thick] (pe)--++(-1.0, 0);
\draw[<-, thick] (pm)--++(-1.5, 0) node[left] {$P_\mathrm{fix}$};
\draw[->, thick] (swing.east)--++(1.5,0);

\node[draw,thick,text width=5cm, below = 0.5cm of swing](pid) {PID Controller
\begin{equation*}
    o(t) = i(t) + \int \dd{t}i(t) + \dv{t}i(t)
\end{equation*}};

\draw[->,thick] (swing.east)++(1,0)|-(pid.east);
\draw[->,thick] (pid.west)-|($(pm)+(-1,0)$) node[anchor=north west]{$-$};

\path (pe)--++(-2.2, -0.3) coordinate (base);
\draw[->] (base)--++(1,0);
\draw[->] (base)--++(0,0.6);
\draw[out=0, in=180] ($(base)+(0.1,0.45)$)--++(0.15,0) to ($(base)+(0.5,0.2)$)--++(0.4,0);

\end{tikzpicture}}\vspace{0.2cm}
\includegraphics[width=.8\columnwidth]{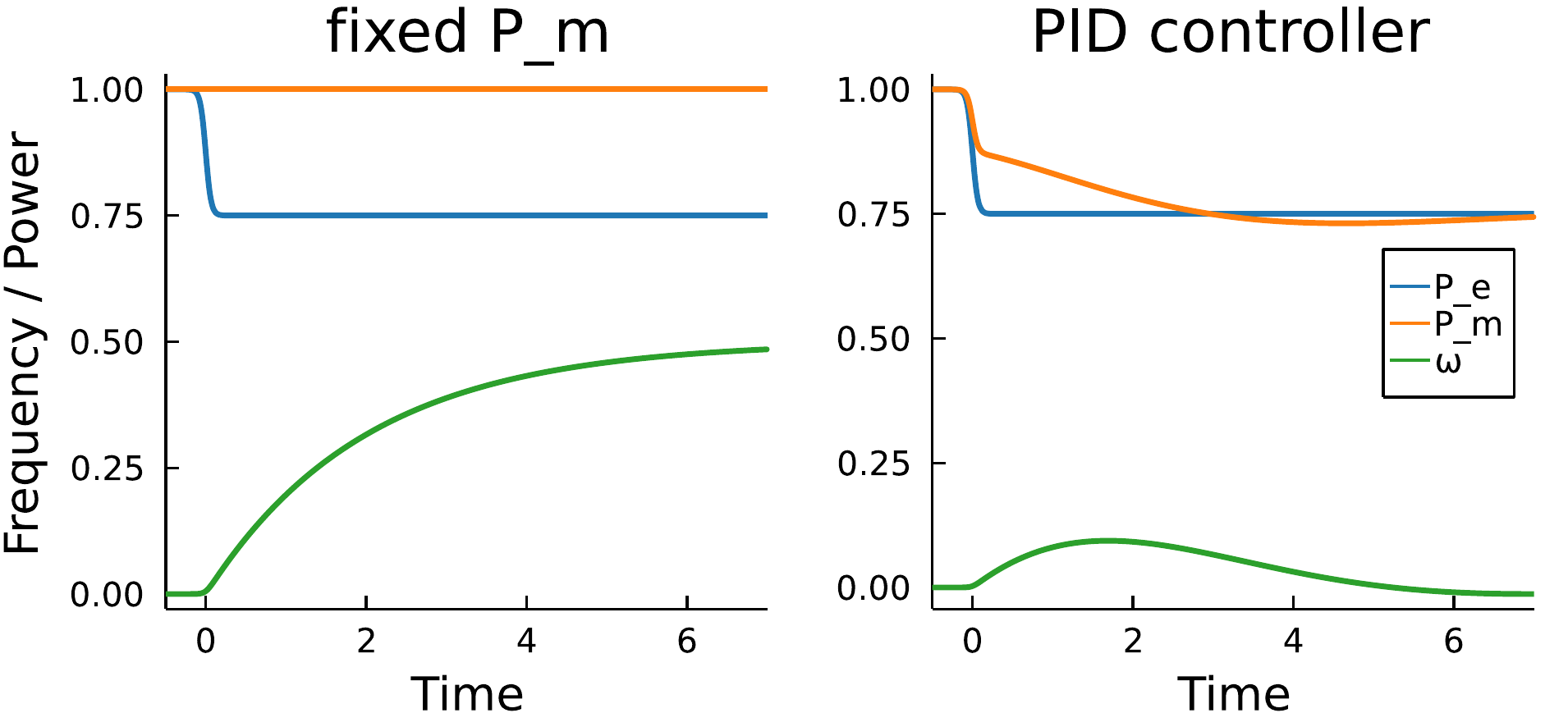}
  \caption{Block diagram and frequency response of the swing node to a sudden drop in electrical power demand with and without PID controller.}\label{fig:pid_example}
\end{figure}
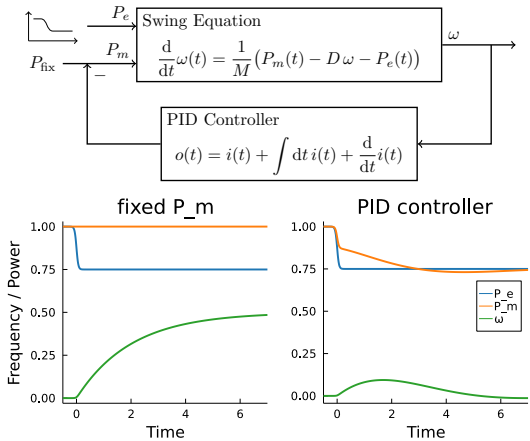

We define a new \texttt{IOSystem} by providing multiple blocks and defining their input-output-relationship.
In this example, we plug the frequency $\omega$ of the swing node into the input of the PID controller.
The output of the PID controller gets plugged into the $P_m$ input of the swing equation.
The resulting closed-loop system is shown in the upper pane of Fig.~\ref{fig:pid_example}.

\begin{minted}{julia}
w_pid = IOSystem([pid.out => swing.P_m,
                  swing.ω => pid.input],
                 [swing, pid])
w_pid = connect_system(w_pid)
\end{minted}

By calling \texttt{connect\_system} the composite system of several building blocks is transformed into a single block with a single set of equations, inputs and outputs.
There are several important steps in this process:
\begin{itemize}
    \item Substitution of input variables with the connected outputs, i.\,e. $P_m$ in the swing equation gets replaced by $1 - pid$.
    \item Substitution of differentials in the right-hand side of the equation by repeated substitution of known terms inside the differential, symbolic expansion of known differentials and substitution of known differentials, i.\,e. $\dv{t}i$ inside the controller gets replaced by $\dv{t}\omega$ gets replaced by $\frac{1}{M}(P_m - D\,\omega - P_e)$.
    \item Reduction of explicit algebraic states which are not marked as outputs of the system. This is done by substituting each occurrence with their right-hand side. The algebraic equation can be removed afterwards.
\end{itemize}
Besides those transformations, there are some additional steps regarding namespacing and namespace promotions of subsystems.

Once the complex system is reduced to a single block the next important step is function building.
Up to this point, all expressions have been symbolic representations of equations rather than Julia functions.
Using the excellent transformation tools from~\texttt{ModelingToolkit.jl} \cite{ma2021modelingtoolkit} and \texttt{Symbolics.jl}, we can generate high performant Julia functions from those equations. Each output and internal state is represented by a state $x_i$ in the state vector $\vb{x}$. The generated function takes an input vector $\vb{i}$ and parameter vector $\vb{p}$ and represent the dynamical system in mass matrix form
\begin{equation}
\vb{M}\cdot\dv{t}\vb{x} = f(\vb{x}, \vb{i}, \vb{p}, t)
\end{equation}
where the (diagonal) mass matrix $\vb{M}$ is also generated by \texttt{BlockSystems.jl}. Here, implicit algebraic constraints (\ref{eq:implicit_constraints}) correspond to zero diagonal entries in the mass matrix. The system can be solved using the solvers from \texttt{DifferentialEquations.jl}. The resulting time series in response to a sudden drop of requested electrical power is shown in Fig.~\ref{fig:pid_example}. The full code for the example can be found in our GitHub repository \cite{github_repo}.

\subsection{NetworkDynamics.jl}
While equation-based modeling with block diagrams is very convenient for the various subcomponents connected to the buses, there are some drawbacks to this technique. Working with symbolic representations of equations is very flexible but does not scale well for very big systems with potentially thousands of buses.

At the network level, we do not need this kind of flexibility because all connections between the components follow strict rules: each node dynamic sees the sum of incoming complex currents on the connected lines as input and defines differential or algebraic equations for the complex node voltage. Lines on the other hand see the voltages of connected nodes as input and calculate the complex currents for both ends.
This kind of modeling is based on the assumption, that the nodal voltage and currents are balanced at all times and the short-term voltage dynamics only react to the currents on the transmission lines. Thus our modeling approach is based on assumptions 1-3 (but not 4 and 5, that restrict the machine models) of \cite{kogler2021normal}.

We can use \texttt{BlockSystems.jl} to generate high performant right-hand side functions for the subcomponents of the system. Those functions are consumed by \texttt{NetworkDynamics.jl}\cite{lindner2021networkdynamics} which constructs right-hand side functions for dynamical systems on networks with heterogeneous components. It automatically constructs an optimized Julia function, ready to be solved with \texttt{DifferentialEquations.jl}.

\subsection{PowerDynamics.jl}
While not being used for modeling the example system in this paper, we also need to mention \texttt{PowerDynamics.jl} as part of our software stack. \texttt{PowerDynamics.jl}\cite{plietzsch2021powerdynamics} is a library of widely used line and generator models as well as fault scenarios from power system theory. Built on top of \texttt{NetworkDynamics.jl} it guarantees efficient implementation and provides a range of convenience functions for plotting and evaluation of the solution, thus leveraging Julia's capabilities for highly efficient numerical simulations. Due to its computational efficiency, it enables probabilistic analyses of power systems, which require a large number of dynamical simulations~\cite{liemann2020probabilistic}. The new block-based modeling we present here will become part of the \texttt{PowerDynamics.jl} in the foreseeable future.

\subsection{ProBeTune.jl}
 \emph{Probabilistic Behavioral Tuning (ProBeTune)} is one among many possible complex optimization problems which can be realized using our software stack. In this publication, we will only use ProBeTune to demonstrate the capabilities of our approach. Hence we refer the interested reader to the original publication \cite{hellmann2021probetune}, which contains the full mathematical background, and only give a short introduction here.

To deal with the varying layers of complexities in power grids it would often be desirable to tune complex systems to behave like a simpler system, a so-called specification. In practical applications, however, it is usually not necessary that such a specification is met precisely. Instead, it is sufficient that the system behaves as \emph{close} to the specification as possible. With ProBeTune Hellmann et. al. \cite{hellmann2021probetune} introduced a method that is used to optimize the internal parameters of a system $p$ such that its dynamics will behave close to a specification. For this purpose, a distance measure between the system and specification is introduced. 
This so-called behavioral distance $d$ is the expected distance between the system and the specification output. In \cite{hellmann2021probetune} it is shown that $d$ can be approximated efficiently by solving a joint optimization problem in $p$ and several copies of the specification parametrized by $q_i$. We refer the reader to \cite{hellmann2021probetune} for details.
The key idea here is that it is not necessary to tune the system towards a particular parametrization of the specification but rather to achieve a certain dynamic behavior.
Julia's differentiable programming capabilities, which enable the use of gradient descent-based optimization methods in combination with differential equations solvers, and the machine learning algorithms of \texttt{DiffEqFlux.jl} \cite{rackauckas2019diffeqflux} allow problems like this to be smoothly solved. A simple interface for the ProBeTune method is implemented in the \texttt{ProBeTune.jl} package.
    
\section{Example: Nordic 5-Bus Test Case}
\label{sec:nordic5}
Transient stability issues of the Nordic electricity grid, due to low inertia, are a well-known problem especially in the summer months when the share of renewable energy sources (RES) is high \cite{inertia_challenge}. As this problem will become more dominant the Nordic Transmission System Operators (TSOs) have introduced a new grid service, the Fast Frequency Reserve (FFR). FFR is designed to assist the already existing Frequency Containment Reserve for Disturbances (FCR-D) in the first 30s after a fault occurred. Markets for FFR have been introduced by all Nordic TSOs in 2020 as promised in \cite{inertia_challenge}.

Recently, the concept of the "Dynamic Virtual Power Plant" (DVPP) has been proposed in \cite{dvpp} as an extension to the "Virtual Power Plant" (VPP). The VPP only coordinates power balance at the minute scale in contrast to the DVPP which includes dynamic phenomena and aids in the dynamical stability of the grid. The main idea of the paper is that RES have to participate in all grid services, including FFR and FCR-D. As in the case of static VPPs, small suppliers of RES are aggregated to aid in ancillary grid services which they would not have been able to supply by themselves. In a companion publication \cite{hydro_and_wind}, the authors presented a method that combines hydro and wind power, allowing their DVPP to fulfill the demands for FCR-D. Their approach solves the instabilities which are related to the "minimum phase zero" of the hydro governors by utilizing the wind turbines as a source for FFR on very short time scales to facilitate the undershoot period of the hydro plants.

However, much research is needed to make this approach more general, since the authors of \cite{hydro_and_wind} have only explored a single scenario, the disconnection of a 1400 MW importing DC link. In addition, it is not possible to include other energy sources into their DVPP. Yet their approach and test case are exciting and ProBeTune offers the perfect framework to handle these types of tasks. The FCR-D guidelines, or any other definition for a DVPP, can be used to define the specification. The system parameters are optimized to behave closest to the specification for a large number of possible scenarios.

\begin{figure}[h]
    \centering
    \includegraphics[width=0.5\columnwidth]{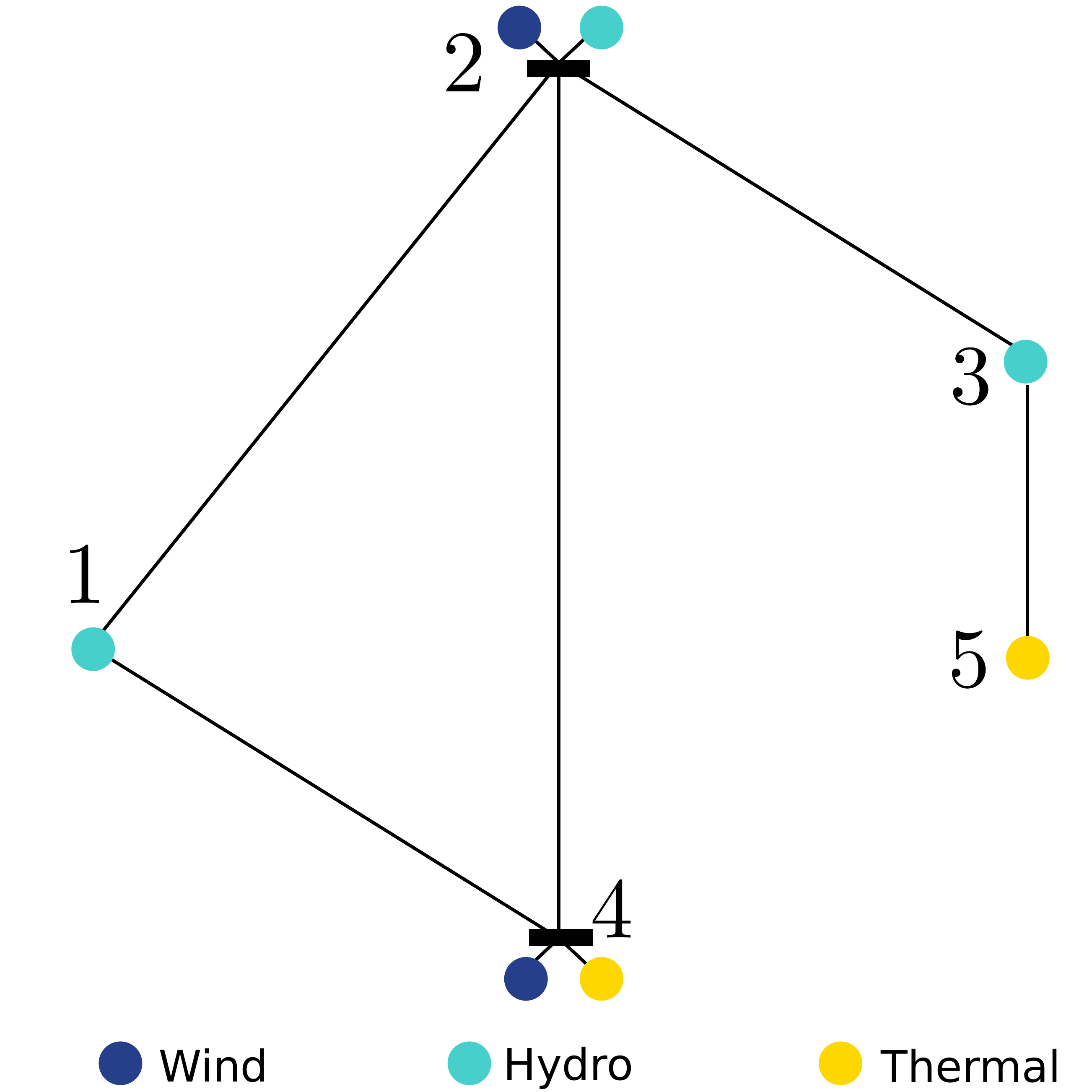}
    \caption{Network structure of the Nordic 5-Bus system. Each Bus consist of a load, a proportional controller and the additional energy sources as indicated.  \label{fig:network}}
\end{figure}

The Nordic 5-bus test system (Nordic5), designed by the authors of \cite{hydro_and_wind}, was implemented using \texttt{BlockSystems.jl} and \texttt{NetworkDynamics.jl}. The nodal dynamics were implemented using \texttt{BlockSystems.jl} by following the Block Diagrams given in the original publication \cite{hydro_and_wind} and adapting them accordingly to our problem.
The Nordic5 setup includes governors, exciters, machines, wind turbines, loads and different types of controllers and therefore already shows many relevant layers of complexity that are needed in realistic power system modeling and are realizable using our software stack. The network structure is shown in Fig.~\ref{fig:network}. Each node consists of a load, a proportional controller and additional energy sources for each node as indicated in Fig.~\ref{fig:network}. The complete modeling of the Nordic5 system can be found in our GitHub repository \cite{github_repo}.

\subsection{Experimental Setup}
For the tuning process, we have fixed the parameters which were given in the publication \cite{hydro_and_wind} while we have assumed that the controllers at the machines, as well as their parameters are adaptable.

Each node in the Nordic5 system contains a machine, whose shaft sets the frequency dynamics. We have added a proportional controller $D_i \omega_i$ to each shaft, resulting in the following frequency dynamics $\omega_i$ for node $i$:

    \begin{align}
        \frac{d\omega_i}{dt} = \frac{1}{2H_i}(P_{m,i} - P_{e,i} - D_i \omega_i)
    \end{align}
where $H_i$ is the inertia constant. $P_{m,i}$, and $ P_{e,i}$ are the mechanical and electrical power respectively and are given by the corresponding differential equations of the machine models, governors and exciters. For example, the thermal machine at bus 5 in Fig.~\ref{fig:network} is modeled using a fifth-order machine model, an exciter and a power system stabilizer.

As a first proof of concept, we will optimize the proportional gain $D_i$ at each machine in the system using \texttt{ProBeTune.jl}. The specification is given by a network of swing equations as the nodal dynamics and with the same topology as the Nordic5 system (Fig.~\ref{fig:network}), including a proportional controller and loads at all buses. The reduction of complex nodal dynamics to a swing equation with a damping term $D_i$ is motivated by the fact that many processes, such as the damper winding effects, constant excitation and damping by loads can as a good approximation be absorbed in the damping term of the swing equation \cite{sauer_pai, anderson_fouad}. Tuning a complex system such that it behaves like a network of simple swing equations can be useful, since the latter are far better understood and also allow for analytical statements, as for example in \cite{doerfler_2012}, that are no longer applicable to much more detailed engineering models.

The output metric $\Delta o$ of system and specification, given in equation \eqref{eq:loss}, is based on a frequency metric that compares the difference between the frequency trajectories of the system and specification after a perturbation, divided by the number of all perturbations $N$:
\begin{align}
    \Delta o = \frac{1}{N} \sum_{j=1}^{N} \sum_{5}^{i=1} \sum_{t} \left(\omega_{i, \mathrm{sys}}^j(t) - \omega_{i, \mathrm{spec}}^j(t)\right)^2, \label{eq:loss}
\end{align}
where $j$ and $t$ run over all perturbation scenarios and 100 uniformly distributed time points in the time series respectively. The frequency trajectory at bus $i$ of either the system or the specification are given by $\omega_i(t)$. The output metric is used to calculate the behavioral distance as defined in \cite{hellmann2021probetune}.

The different perturbation scenarios $j$ are given by randomly choosing a bus $i$ to be perturbed and sampling a normally distributed power perturbation $\Delta P_i$ which changes the demand on the load of bus $i$. The frequency metric \eqref{eq:loss} is minimized by adapting the proportional gains $D_i$ at all nodes. 

The initial gains of the system have been uniformly drawn from an interval of $[0, 1]$ while the gains of the specification were drawn from $[0, 5]$.
As the first step, we calculate the behavioral distance $d$ of the system and specification as defined in \cite{hellmann2021probetune} to get a measure of how similar system and specification are already. Then we tune the parameters of the system and specification until the frequency metric converges. Finally, we calculate the behavioral distance again with the tuned parameters.
Our optimization is performed on 10 different perturbation scenarios using the ADAM optimizer \cite{kingma2014adam}.

\subsection{Results}
The initial behavioral distance between system and specification is 13.36 which indicates that the dynamical behavior of the both is not very similar. 

After the joint optimization of the system and specification parameters has converged we calculate the behavioral distance again. The final behavioral distance is $0.005$ and a reduction of the behavioral distance of a factor $\sim 2684$ can be reported. This means that the frequency dynamics of the complex Nordic5 system can be described by swing equations and shows that our approach was successful.

\begin{figure}%[h]
    \centering
    \includegraphics[width=.9\columnwidth]{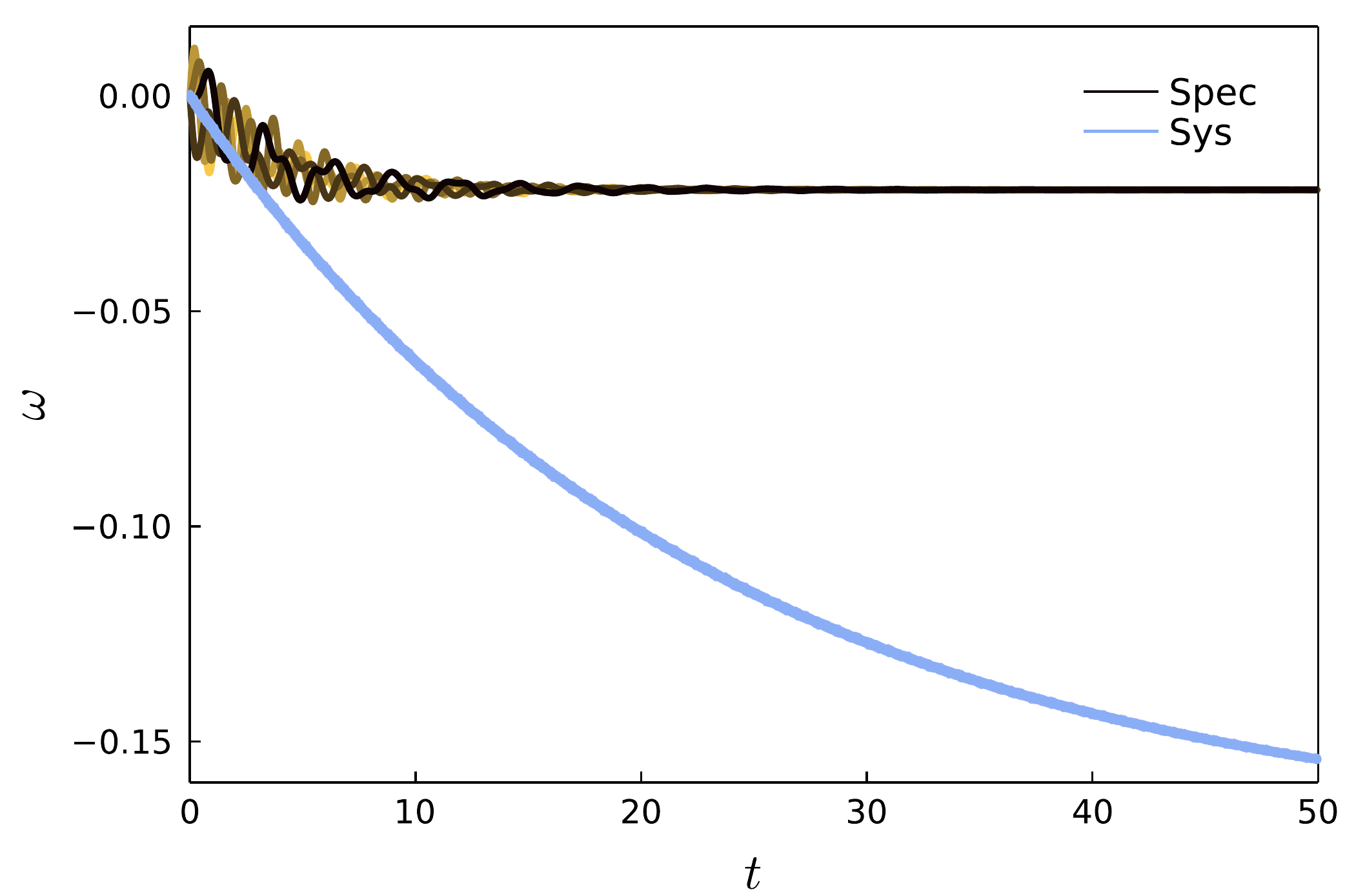}
    \caption{Frequency transients of the initial system and specification after a power perturbation on bus 4. The system and specification do not inherit the same fixed point. \label{fig:untuned}}
\end{figure}

\begin{figure}%[h]
    \centering
    \includegraphics[width=.9\columnwidth]{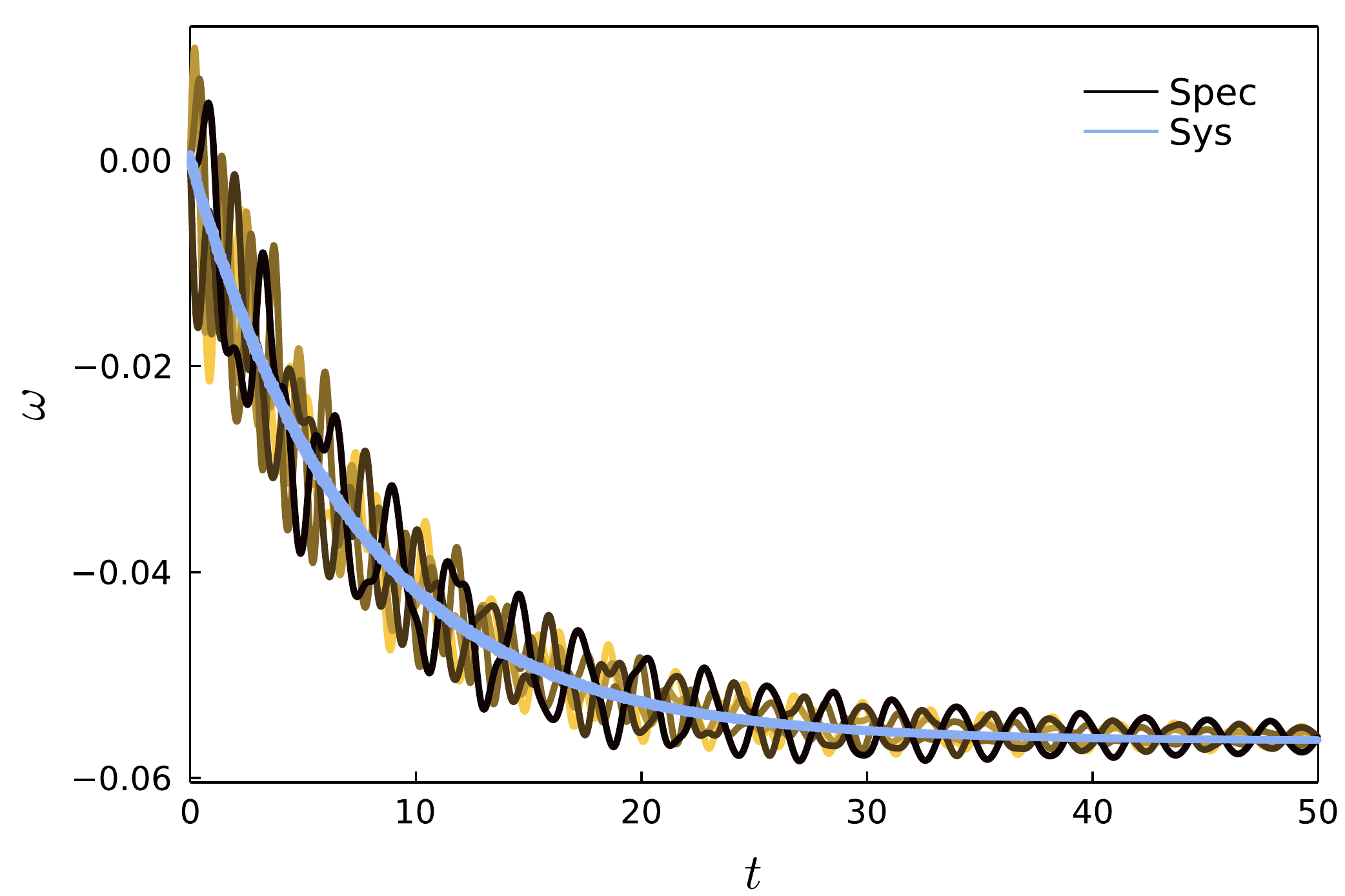}
    \caption{Frequency transients of the tuned system and specification after a power perturbation on bus 4. The system and specification inherit the same fixed point. \label{fig:tuned}}
\end{figure}

Fig.~\ref{fig:untuned} and Fig.~\ref{fig:tuned} show the frequency transients of the system and specification after the same power perturbation on bus 4. Fig.~\ref{fig:untuned} shows the initial system with random proportional gains while Fig.~\ref{fig:tuned} shows the transients after the tuning process. The frequency transients of the nodes in the Nordic5 system overlap so well that they become indistinguishable in Fig.~\ref{fig:untuned} and \ref{fig:tuned}.

In Fig.~\ref{fig:untuned} the system and specification run off to different fixed points after the perturbation. The specification settles on a fixed point closer to the synchronous frequency and does so quicker, which is to be expected since the specification was initialized with higher proportional gains.

Fig.~\ref{fig:tuned} shows the tuned system and specification after the same perturbation as in the previous Fig.~\ref{fig:untuned}. After the training both, system and specification, settle on the same fixed point, within the same time frame. The systems are only distinguishable by the higher harmonic oscillations of the specification. 

While the specification allows for more oscillations, as the nodes in the Nordic5 system contain various controllers to prevent this, the results are still meaningful as they show that we can reduce very complex dynamics onto a simpler model. 
These results should be treated as a proof of concept to demonstrate the software stack, to apply ProBeTune properly we have to use more sophisticated specifications and tuning processes.
%highlight the potentials our approach holds and opens possibilities for more sophisticated specifications and tuning processes.

The next step is to use controllers which can restore the synchronous frequency, like a PID controller, or to reduce the Nordic5 system onto a single swing equation. Finally, the ultimate goal is to use the design target for the power output, given in \cite{hydro_and_wind}, as the specification to define a proper DVPP.
    
%\newpage
\section{Outlook}
We have presented our purely Julia-based software stack, which allows us to build modular equation-based networks by using \texttt{NetworkDynamics.jl} and \texttt{BlockSystems.jl}. We have shown that we can use this stack to implement and simulate a highly relevant test case, presented in Sec.~\ref{sec:nordic5}, with complicated, heterogeneous dynamics on the buses. Finally, using Julia's capabilities for automatic differentiation and machine learning we have tuned the control parameters of this complex system to behave like a network of swing equations using \texttt{ProBeTune.jl}. This complex example showed the capabilities of our software and may serve as a starting point for more elaborate tuning processes and control schemes, such as the implementation of a “Dynamical Virtual Power Plant“ \cite{dvpp}. Future work on our software stack will include the implementation of GPU acceleration and Jacobian-free solver methods to further reduce simulation times and enable Monte Carlo simulations of fault scenarios for realistically sized power grids.

\section*{Code availability}

All code to reproduce the results and figures of this paper is available at the DOI \url{https://zenodo.org/record/5744264} or at the GitHub repository \url{https://github.com/PIK-ICoNe/Open-Power-Simulation-Conference-2021-Paper}.

\section*{Acknowledgment}

The authors acknowledge the support of BMBF, CoNDyNet2 FK. 03EK3055A and the Deutsche Forschungsgemeinschaft (DFG, German Research Foundation) – KU 837/39-1 / RA 516/13-1 \& HE 6698/4-1.
A. Büttner acknowledges support by the German Academic Scholarship Foundation.
Mr. Lindner greatly acknowledges support by the Berlin International Graduate School in Model and Simulation based Research (BIMoS) of the TU Berlin.
All authors gratefully acknowledge the European Regional Development Fund (ERDF), the German Federal Ministry of Education and Research and the Land Brandenburg for supporting this project by providing resources on the high-performance computer system at the Potsdam Institute for Climate Impact Research.

\bibliographystyle{IEEEtran}
\bibliography{main}

\end{document}